# Observation of plateau-like magnetoresistance in twisted Fe$_3$GeTe$_2$/Fe$_3$GeTe$_2$ junction

Junghyun Kim[1,2,3], Suhan Son[1,2,3], Matthew. J. Coak[1,2,4], Inho Hwang[1,2,3], Youjin Lee[1,2,3], Kaixuan Zhang[1,2,3*], and Je-Geun Park[1,2,3]

[1]Center for Correlated Electron Systems, Institute for Basic Science, Seoul 08826, South Korea
[2]Department of Physics and Astronomy, Seoul National University, Seoul 08826, South Korea
[3]Center for Quantum Materials, Seoul National University, Seoul 08826, South Korea
[4]Department of Physics, University of Warwick, Coventry CV4 7AL, United Kingdom

Corresponding Author: kxzhang@snu.ac.kr

Controlling the stacking of van der Waals (vdW) materials is found to produce exciting new findings, since hetero- or homo- structures have added the diverse possibility of assembly and manipulated functionalities. However, so far, the homostructure with a twisted angle based on the magnetic vdW materials remains unexplored. Here, we achieved a twisted magnetic vdW Fe$_3$GeTe$_2$/Fe$_3$GeTe$_2$ junction with broken crystalline symmetry. A clean and metallic vdW junction is evidenced by the temperature-dependent resistance and the linear I-V curve. Unlike the pristine FGT, a plateau-like magnetoresistance (PMR) is observed in the magnetotransport of our homojunction due to the antiparallel magnetic configurations of the two FGT layers. The PMR ratio is found to be ~0.05% and gets monotonically enhanced as temperature decreases like a metallic giant magnetoresistance (GMR). Such a tiny PMR ratio is at least three orders of magnitude smaller than the tunneling magnetoresistance (TMR) ratio, justifying our clean metallic junction without a spacer. Our findings demonstrate the feasibility of the controllable homostructure and shed light on future spintronics using magnetic vdW materials.

**Keywords**: magnetic van der Waals materials, magnetic homojunction, plateau-like magnetoresistance, Fe$_3$GeTe$_2$

## I. INTRODUCTION

After the success of graphene and two-dimensional (2D) van der Waals (vdW) materials, magnetic vdW materials have been recently attracting enormous attention [1] [2]: e.g., insulating antiferromagnets NiPS$_3$ [3] and FePS$_3$ [4], insulating ferromagnet CrI$_3$ [5], Cr$_2$GeTe$_6$ [6], VI$_3$ [7] and metallic ferromagnet Fe$_3$GeTe$_2$ [8]. Among these magnetic vdW materials, the ferromagnetic materials are particularly interesting from the viewpoint of spintronics. They have already been shown to have spin filter effect [9] and tunneling magnetoresistance (TMR) [10] [11]. Besides, the possible assembly of various layered materials, i.e., vdW heterostructures herald more opportunities and flexibilities in the emerging novel spintronics [12] [13].

Metallic ferromagnetic (FM) vdW materials are useful building blocks for new heterostructures due to their compatibilities in spintronic applications [14]. And Fe$_3$GeTe$_2$ (FGT) is the only known example of vdW FM metal (FMM) with topological band structures. It exhibits a large perpendicular magnetic anisotropy with a clear dependence on temperatures and sample thicknesses [15] [16]. Nanometers-thick FGT is a typical hard ferromagnet, exhibiting significant anomalous Hall effect (AHE) with a near-square-shaped hysteresis loop. Furthermore, the recent studies on the tri-layer FGT/Graphite/FGT and FGT/h-BN/FGT heterostructure were reported [17] [18]. Such tri-layer FMM/spacer/FMM constitutes the representative spin-valve model for the well-known giant magnetoresistance (GMR) and tunneling magnetoresistance (TMR) effect [19] [20], where the spacer is a non-magnetic metallic and insulating layer for the GMR and TMR, respectively. Conventionally, GMR and TMR devices were fabricated with magnetic thin films by the magnetron sputtering or the epitaxial growth. Thereby, it naturally limits the stacking flexibility and diversity of the hetero- or homo- structure, and thus hinders further exciting developments in possible controllable homostructures.

In the present work, we achieved a twisted FMM/FMM structure using FGT nanoflake as a model system. Without a spacer between the two FGT nanoflakes, a FGT/FGT structure with a clean interface and a metallic vdW junction was demonstrated. Furthermore, a plateau-like magnetoresistance behavior is found due to the antiparallel magnetic configurations of the two FGT nanoflakes.

## II. EXPERIMENTAL

FGT single crystals were grown by the chemical vapor transport method with iodine as the transport agent. High purity Fe (99.998%), Ge (99.999%), and Te (99.999%) powders were mixed in a stoichiometric ratio 3:1:2. The mixed powder was sealed into an evacuated quartz glass ampoule, which was then placed in a two-zone furnace with a temperature gradient of 750 to 650℃ for one week.

The nanometers-thick FGT flakes were mechanically exfoliated from the as-synthesized crystals onto the 285 nm-thick $SiO_2$/Si substrates. Starting from the same lattice orientation of the selected nanoflake, we stacked a misoriented structure. Using the dry-transfer method, we took apart half of the FGT nanoflake, then picked it up using a polymer stamp. The picked FGT flake was first rotated by a specific angle and then dropped down on the remaining half of the nanoflake. Finally, the whole entity of the twisted FGT/FGT structure was dry transferred onto the pre-patterned 30/5 nm Au/Ti electrodes on the $SiO_2$/Si substrates. All the fabrication process was performed inside the glove box (Oxygen < 0.6 ppm, $H_2O$ < 0.2 ppm) to prevent the surface oxidation of FGT. The transport measurement was conducted using a commercial system (PPMS-9, Quantum Design).

## III. RESULTS AND DISCUSSION

In the FGT/FGT structure shown in Fig. 1 (a), a misoriented vdW lattice exists in the interface between Top- and Bottom-FGT. Figure 1(b) shows the optical image of the constructed FGT/FGT sample, where the twisted angle $\theta$ is about 87°. In our fabrication process, we could get a freely controlled twisted-stacked structure. Both Top- and Bottom-FGT should have the same properties (e.g., resistance) as they came from the identical nanoflake.

The clean interface is essential to achieve a perfect vdW homojunction. The vertical transport property through the junction is suitable for verifying the clean interface because it largely depends on the density of states near the Fermi level and the effective barrier. We were able to prevent an insulating oxidized FGT layer in the junction, by exfoliating FGT nanoflake within the inert gas. Figure 1(c) shows the metallic behavior of the junction resistance, where the resistance $R$ reduces as the temperature $T$ decreases. The $R$-$T$ behavior of the junction is identical to that of pristine FGT, with a high magnetic transition temperature and a resistance minimum [8]. As shown in Fig. 1(d), the perfectly linear *I-V* curves for both the junction and the single Top-FGT at 2 K confirm the clean metallic interface in the vdW homojunction. Note that, though the interface is metallic, the top surface of Top-FGT might be slightly oxidized due to exposure to the air while transferring the sample from the glove box to the measurement cryostat. The oxidized top surface will reduce the effective thickness and induce possible domain pinning in Top-FGT, resulting in a larger $H_C$ compared to that of Bottom-FGT [16].

Next, we investigated the junction magnetoresistance as a function of a perpendicular magnetic field, with current flowing from Top-FGT to Bottom-FGT. The FGT/FGT structure was expected to behave as a single FGT nanoflake due to the absence of the spacer. On the contrary, the junction magnetoresistance turns out to depend mainly on the relative magnetization direction of the Bottom-FGT and Top-FGT. Here, we observed the resistance plateau with the steep jump and drop as well as the linear magnetoresistance below $T_C$ (see Fig. 2(b) ~ (e)). Figure 2(a) displays the schematic of the junction magnetoresistance, which is composed of both the single FGT (i.e., dashed line) and the junction (i.e., solid line) contribution. The linear magnetoresistance for single FGT peaks near the coercive field $H_C$, making butterfly-shaped resistance. While the magnetic field $H$ increases, more magnetic moments of the single FGT align along $H$, leading to magnon-suppressed scattering and linearly reduced resistance [21]. Linear positive magnetoresistance in butterfly-shaped region is arising from the enhanced magnon population, as opposed to the magnon suppression for the linear negative magnetoresistance. Such butterfly-shaped magnetoresistance can be arising from the oxidized surface of FGT. It was revealed that the oxidized surface of FGT can generate tilted spins in the samples [22], which will lead to the butterfly-shaped magnetoresistance through the well-known anisotropic magnetoresistance effect and the magnon scattering in ferromagnets.

As shown in Fig. 2(a), the resistance plateau emerges when the antiparallel magnetic configuration develops in the field range of $H_{C, Bottom} < H < H_{C, Top}$. We explain that initially, the magnetic moments point along the down direction for both the Bottom-FGT and Top-FGT. When $H$ increases to the positive $H_{C, Bottom}$, the magnetic moments of Bottom-FGT flip along $H$ while that of Top-FGT remain down against $H$. When $H$ further increases to $H_{C, Top}$, the magnetic moments of Top-FGT also flip along the $H$ as that of the Bottom-FGT. Therefore, the antiparallel magnetic configuration develops in $H_{C, Bottom} < H < H_{C, Top}$ in an increasing magnetic field. Since the spin-dependent scattering is more significant for an antiparallel alignment, a high resistance plateau emerges. On the other hand, a decreasing magnetic field also produces a similar antiparallel magnetization state with a corresponding resistance plateau seen in the negative magnetic field range. Here, the higher resistance plateau for the antiparallel than the parallel magnetic configurations in such two metallic FGTs resembles a GMR structure with a metallic spacer.

In our results, the symmetrical plateau-like magnetoresistance (PMR) was observed at 100 and 50 K (see Fig. 2(b) and (c)), which is similar to the typical GMR and TMR phenomena. Although the PMR also appears in the lower temperatures (see Fig. 2(d) and (e)), the junction MR shows slightly asymmetric behavior. Accordingly, such asymmetric behavior also exists in the single Top-FGT at low temperatures like 20 K, as shown in Fig. 3. The purple area box covers from the left coercive field $H_{C, L}$ (i.e., left peak) to the right coercive field $H_{C, R}$ (i.e., right peak). The asymmetric behavior $H_{C, L} > H_{C, R}$ is visible in the data taken at 20 K. Due to the asymmetric coercivity, we think that the $H_{C, L}$ difference between the Bottom-FGT and Top-FGT is too small to form a well-defined antiparallel magnetic state. Therefore, the left high- resistance plateau is absent at 10 and 2 K, as seen in Fig. 2(d) and (e). This asymmetricity could also arise from the anisotropic interfacial interaction or the domain pinning effect of Top-FGT. It becomes a dominant factor in reducing the thermal activation at low temperatures, whose detailed study is warranted in the future.

The PMR ratio possesses a reasonable value for the metallic vdW FGT/FGT junction. The PMR ratio is defined as $[R_{AP} - R_P]/R_P$, where $R_{AP}$ and $R_P$ represent the resistance values of the antiparallel and parallel configuration, respectively. As summarized in Fig. 4 (b), the PMR ratio is ~0.05%, which is similar to that of conventional GMR devices (0.1~15%) [19], but three orders of magnitude smaller than that of typical TMR (70~600%) devices [23]. The small value of PMR ratio in our device implies the metallic interface with the absence of an insulating oxidized FGT spacer, which belongs to the range of the GMR ratio.

The temperature dependence of the PMR ratio shows a monotonic increase by ~0.025% from 100 to 2 K, as shown in Fig. 4(b). The PMR effect is suppressed at higher temperatures, which can be explained similarly to the reported GMR spin-valve effect [25]. As the temperature increases, the spin polarization decreases. Meanwhile, the expected momentum transfers between the spin-up and spin-down electrons will occur in the ferromagnetic layers. At the interface, the electron-magnon scattering is anticipated to give rise to spin mixing through spin flipping, which suppresses the relative contribution of the spin-dependent scattering [25]. Therefore, the PMR ratio is reduced at higher temperatures. A passing remark, Fig. 4(b) inset shows that the coercive field $H_{C,\,Top}$ also increases as the temperature decreases. Previous works revealed that the perpendicular magnetic anisotropy of the FGT increases with decreasing temperature [16]. Moreover, the magnon population is substantially suppressed at lower temperatures [26]. Both factors collectively lead to the enhanced $H_{C,\,Top}$ and $H_{C,\,Bottom}$ in decreasing temperatures.

Here we would like to discuss the specialty of our spin valve homojunction. In principle, two ferromagnetic metals will not work because they will be strongly coupled with no spacer. However, the vdW gap in the twisted FGT/FGT homojunction could decouple two ferromagnetic layers by much weakening the interlayer coupling [24]. Such weak interlayer coupling between two ferromagnetic layers could help to form different coercivity and stabilize the antiparallel magnetic states, eventually generating the PMR effect. Moreover, the steep resistance changes of plateaus support the absence of the exchange field in the vdW homojunction. Thus, whether the PMR effect is significant enough to be observed depends on a few ingredients including the large spin polarization, weak interlayer magnetic coupling and strong stability of the antiparallel magnetic state for the device, etc.

The vdW junction of the FGT/FGT structure can apply to other possible homojunction. And a similar by-layer effect was also reported in other systems like $NbSe_2/NbSe_2$ homostructure [27]. Moreover, the twisted interface holds an exciting possibility to develop a crystalline symmetry broken bilayer FGT in the junction. We expect to tailor the interface function by controlling the angle in further study, which may promise the 2D twistronics [28] using magnetic vdW materials.

### IV. CONCLUSIONS
In summary, we achieved a FGT/FGT vdW homojunction structure with the twisted interface. The metallic behavior of the junction resistance confirms the clean interface. In such a homojunction of FGT nanoflake, the plateau-like magnetoresistance (PMR) was observed below $T_C$ due to the antiparallel magnetization configuration. Unlike the GMR and TMR devices that require the additional separating spacer, the vdW homojunction here stabilizes a resistance plateau without a spacer. Our findings promise the new type of twisted magnetic vdW devices in 2D spintronics.

**ACKNOWLEDGMENTS**

The work at IBS CCES was carried out with financial support from the Institute for Basic Science (IBS) of the Republic of Korea (Grant No. IBS-R009-G1) and CQM was supported by the National Research Foundation of Korea (Grant No. 2020R1A3B2079375).

**DATA AVAILABILITY**

The data that support the findings of this study are available from the corresponding author upon reasonable request.

**Figures**

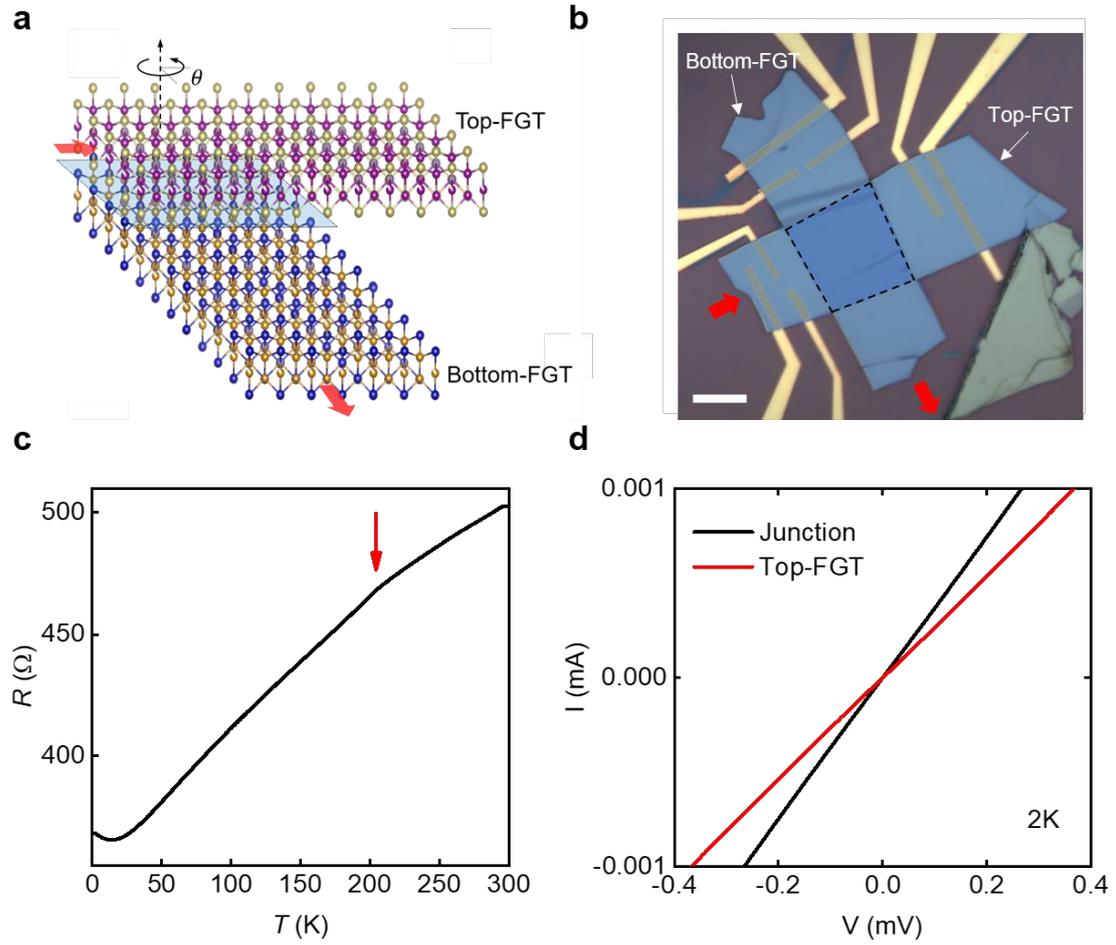

FIG. 1. (a) Schematic of FGT/FGT structure. The orientation of Top-FGT lattice and Bottom-FGT lattice makes an angle of θ. Blue area indicates a twisted interface between Top- and Bottom-FGT of the junction. (b) Optical image of FGT/FGT structure with blue false-color. The inserted scale bar is 10 μm. The dashed line indicates a stacked region where the lattice has a misaligned vdW junction, corresponding to the blue area in (a). Top-FGT was torn out from an original FGT nanoflake and then transferred on the remaining Bottom-FGT. The electrodes are contacted to the bottom of each FGT layer. (c) Junction resistance as a function of temperature at 0 T. The red arrow indicates the Curie temperature of 206 K. (d) I-V characteristics of the junction and Top-FGT resistance at 2 K. Vertical (in-plane) current was adopted in the measurement for the junction (single Top-FGT) resistance.

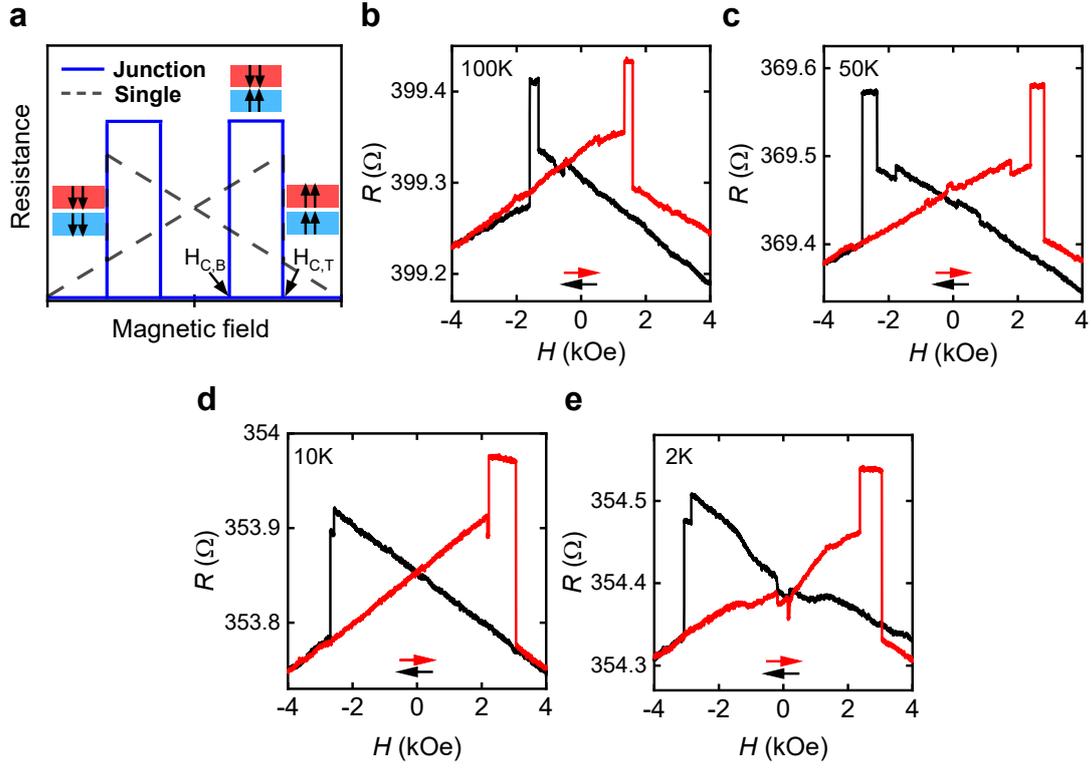

FIG. 2. (a) Schematic of the junction magnetoresistance. Total resistance is composed of both single FGT (dashed line) contribution and junction (solid blue line) contribution. Spin configurations of Top-FGT (red block) and Bottom-FGT (blue block) are shown in each resistance region, according to the parallel and antiparallel states. (b)~(e) Junction magnetoresistance curve with the applied magnetic field swept from –4 to 4 kOe at 100, 50, 10, and 2 K. Below $T_C$, the junction magnetoresistance shows plateaus with a larger value of the resistance than that of the linear magnetoresistance.

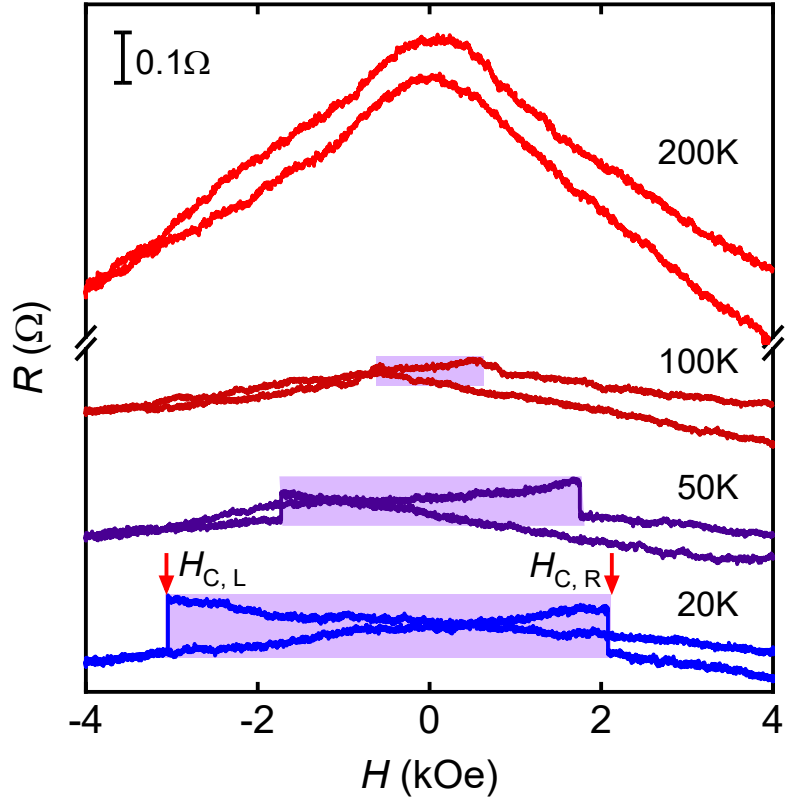

FIG. 3. Magnetoresistance curve measured with in-plane current flowing through the Top-FGT measured at 20, 50, 100, and 200 K. The purple block indicates the butterfly-shaped region covering from the left coercive field $H_{C,L}$ (i.e., left peak) to the right coercive field $H_{C,R}$ (i.e., right peak), where the spin-flip transition occurs. Both $H_{C,L}$ and $H_{C,R}$ increase as temperature decreases. As the temperature is lowered, the asymmetric behavior becomes larger around the zero magnetic fields as a center point.

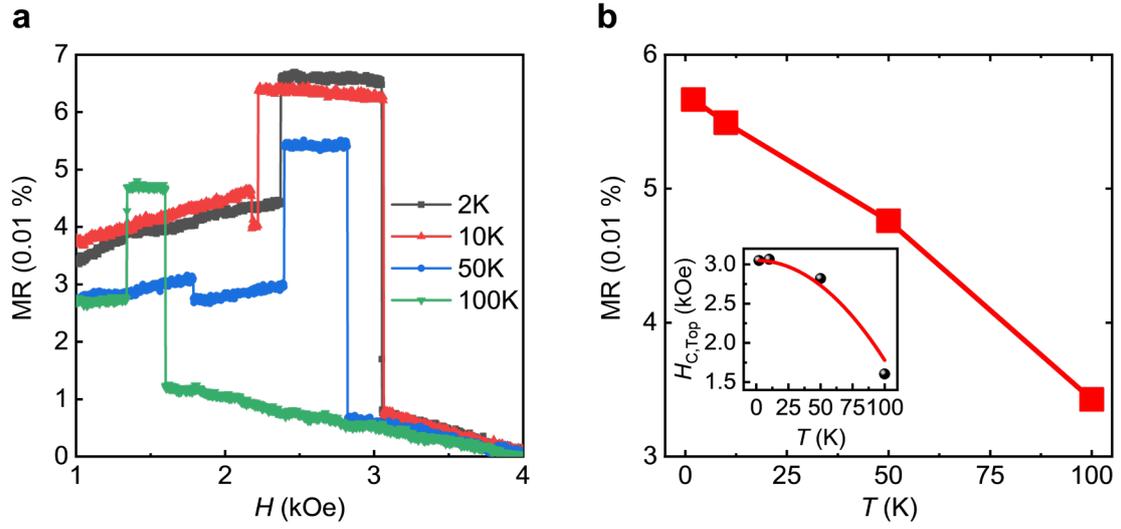

FIG. 4. (a) Magnetoresistance (MR) ratio in the magnetic field from 1 to 4 kOe measured at 2, 10, 50, and 100 K. The MR ratio is calculated $[R(H) - R_{\max}]/R_{\max}$, where $R_{\max}$ is the resistance at 0.4 T. (b) Plot for PMR ratio from 2 to 100 K. The PMR ratio is the value of difference at $H_{\text{C,Top}}$. The linearly decreasing behavior is shown as the temperature increases. The inset shows a plot for the temperature-dependent $H_{\text{C,Top}}$, where the solid red line is a guide to eyes.